\documentclass[%
 aip,
 rsi,
 amsmath,amssymb, 
 reprint,%
]{revtex4-1}

\usepackage{graphicx}
\usepackage{graphics}
\usepackage{dcolumn}
\usepackage{bm}

\usepackage[utf8]{inputenc}
\usepackage[T1]{fontenc}
\usepackage{mathptmx}
\usepackage{etoolbox}
\usepackage{xcolor}
\usepackage{comment}
\usepackage[thinlines]{easytable}
\usepackage{subfigure}

\graphicspath{{Figures/}}

\usepackage{amsmath}
\usepackage{hyperref}
\usepackage{gensymb}
\usepackage{multirow}

\renewcommand{\arraystretch}{1.4}
\makeatletter
\def\@email#1#2{%
 \endgroup
 \patchcmd{\titleblock@produce}
  {\frontmatter@RRAPformat}
  {\frontmatter@RRAPformat{\produce@RRAP{*#1\href{mailto:#2}{#2}}}\frontmatter@RRAPformat}
  {}{}
}%

\newcommand{\correction}{} 

\makeatother
\begin{document}

\preprint{AIP/123-QED}

\title[A Magnetic Eye-Tracker]{An innovative eye-tracker. Main features and demonstrative tests}
\author{Lorenzo Bellizzi}
\affiliation{DIISM University of Siena, Via Roma 56, 53100 Siena (Italy)}

\author{Giuseppe Bevilacqua}
\affiliation{DSFTA University of Siena, Via Roma 56, 53100 Siena (Italy)}

\author{Valerio Biancalana}
\email{valerio.biancalana@unisi.it}
\affiliation{DSFTA University of Siena, Via Roma 56, 53100 Siena (Italy)}

\author{Mario Carucci}
\affiliation{DSMCN  University of Siena, UOC Otorinolaringoiatria, Viale Bracci 16,  53100 Siena, (Italy)}

\author{Roberto Cecchi}
\affiliation{DSFTA University of Siena, Via Roma 56, 53100 Siena (Italy)}

\author{Piero Chessa}
\affiliation{{Dept. of Physics "E.Fermi", University of Pisa, Largo Pontecorvo 3, 56127 Pisa, Italy}}

\author{Aniello Donniacuo}
\affiliation{DSMCN  University of Siena, UOC Otorinolaringoiatria, Viale Bracci 16,  53100 Siena, (Italy)}

\author{Marco Mandalà}
\affiliation{DSMCN  University of Siena, UOC Otorinolaringoiatria, Viale Bracci 16,  53100 Siena, (Italy)}

\author{Leonardo Stiaccini}
\affiliation{DSFTA University of Siena, Via Roma 56, 53100 Siena (Italy)}

\date{\today}

\begin{abstract}
We present a set of results obtained  with an innovative eye-tracker based on magnetic dipole localization by means of an array of magnetoresistive sensors. The system  tracks both head and eye movements with a high rate (100-200 Sa/s) and in real time. A simple setup is arranged to simulate head and eye motions and to test the tracker performance under realistic conditions. Multimedia material is provided to substantiate and exemplify the results. \correction{A comparison with other available technologies for eye tracking is drawn, discussing advantages (e.g. precision) and disadvantages (e.g. invasivity) of the diverse approaches, the presented method standing out for low cost, robustness and relatively low invasivity.}
\end{abstract}

\maketitle

\section{Introduction}
We develop a magnetic tracker \cite{brevetto} suited to produce detailed information about eye and head movements of a patient. The system is based on an array of magnetoresistive detectors that are fixed to the patient head and measure with a high rate and in multiple assigned locations the magnetic field produced by a small size magnet inserted in a contact lens. 

The analysis of the magnetometric data lets evaluate the position and the orientation of the magnet, as well as the orientation of the detectors with respect to the environmental field, which is assumed substantially homogeneous over the volume of the sensor array. The former estimation lets retrieve the position of the eye with respect to the array of detectors, and the latter lets retrieve the orientation of the patient's head with respect to the ambient. In conclusion, simultaneous eye and head orientations can be estimated, which, beside enabling an absolute information of the glance direction, is of interest in several kinds of medical diagnostics, and particularly for those based on comparison between head and eye motion.

Two previous papers have been published to describe the hardware \cite{biancalana_instrHW_21} of that magnetic tracker, and the approaches developed for data analysis \cite{biancalana_instrSW_21}. The  time and precision performance is discussed in Ref.\onlinecite{biancalana_instrSW_21}, together with reliability (robustness) of the algorithms used in data analysis and other details of the applied strategies. 

With this work, we present the application of the developed instrumentation to track head and eye movements with the help of a very simple hand-actuated system that simulates the eye and the head motion. 

After a summary of the available technologies for eye-tracking (Sec. \ref{sec:stateoftheart}), we present the setup (Sec.\ref{sec:setup}) and the approach used to estimate eye and head rotations around a vertical and a transverse, horizontal direction (Sec. \ref{sec:calculus}). A sample of significant results, accompanied by demonstrative media-data available online, is reported in Sec.\ref{sec:results}.

\section{State of the art}
\label{sec:stateoftheart}

There exists a variety of application fields for eye-tracking, ranging from augmented and virtual reality (including entertainment), robotics, aeronautics, medicine \cite{eggert_no_07, carter_ijpp_20}.
Correspondingly, different levels of precision,  robustness and non-invasivity are required. In addition, the intrusivity  (i.e. the risk that the used instrumentation alters or limits the patient response and the eye motion) is worth of being considered. Different techniques have been proposed that are based on measuring diverse  physical quantities and feature various combinations of the characteristics mentioned here above.

Concerning the tracked quantities, eye-trackers can be divided in two 
groups: some of them measure angular eye position relative to the head, while others measure eye position relative to the surroundings \cite{singh_ijsrp_12}. 

\subsection{Electro-oculography}
The \text{Electro-Oculography} (EOG) is one of the 
oldest methods \cite{kolder_oph_74} to track the movement of eyes with respect to the head. Several 
electrodes are arranged around the eye and record the 
standing corneal retinal potential arising from 
hyperpolarizations and depolarizations existing between the cornea and the retina. This technique is relatively
economical, easy to use and maintain and requires a 
 data acquisition system that detects a signal that is already electric in nature.The EOG can register eyes movements when the eyelids are closed and when the patient is sleeping or not-collaborative. On the other hand, relevant problems arise from the noise generated by blinking and movement of facial muscles and from slow drifts of the necessary DC-coupled, high-gain amplifier \cite{aungsakun_ijps_12}.

\subsection{Infrared camera IR-cam}

Infrared OculoGraphy \cite{kumar_lar_92, aungsakun_ijps_12} (IROG) constitutes another tool to track the eye movements with respect to the head. This technique is based on detecting the infrared light scattered by the frontal surface of the eyeball. 
A fixed  IR light source illuminates the eye and photodetectors collect the  light reflected towards various directions. All the instruments are located on a goggle-shaped frame worn by the patient: such a simple setup makes IROG a minimally invasive method.  
The IROG lets measure eye movements in darkness and has a good spatial and time resolution. Unfortunately, this technique works well for small horizontal and vertical angles. The IR has less noise than EOG but is very sensitive to changes of external light and requires a patient-based recalibration. This implies that the system must be calibrated when applied to different patients, and may produce distorted outputs when the environment illumination changes \cite{chennamma_IJCSE_13}. Signal loss is caused by blinking and the method cannot be applied to sleeping patients.

\subsection{Video-oculography}

The Video-oculography (VOG) method uses a single or multiple cameras to determine the eye orientation using the information obtained from the images captured. The cameras can be fixed to the head (in this case the system measures the angular eye position relative to the head) or to an external frame (in this case the system retrieves the eye position relative to the surroundings). Similarly to IROG, VOG has the advantage of being minimally invasive \cite{kimmel_fbn_12, houben_iovs_06} and the disadvantage of being hindered by blinking and not viable with sleeping patients. A complex hardware is required due to the need of fast-cameras (several hundreds frames per second) and subsequent image analysis with cumbersome data elaboration.

\subsection{Search coil}

The gold standard for high
resolution eye tracking is the magnetic tracking with scleral
search coils (SSC), based on Faraday's induction law \cite{robinson_ieee_63, collewijn_vr_75}. 
A time dependent magnetic field induces an electromotive force on the coil(s) accordingly to their reciprocal orientation.
Alternating field components can be applied with different frequencies along three spatial directions, in such way that a harmonic analysis of the induced voltage provides a simultaneous estimate of the directional cosines, i.e. of the coil orientation with respect to the field generators. The latter can be  fixed either to the head or to an external frame, as to produce relative or absolute angle estimations.

With such a setup this method results quite invasive, particularly due to the wire that leads the induced signal to the measuring instrumentation, but also intrusive for the need of rather bulky field generators
. Thus the VOG approach keeps being a favorite choice whenever it can be applied \cite{zee_on_14}. 

Several projects have been proposed to reduce the SSC invasivity level by dropping that wired connection. 
Reulen and Bakker  suggested to
measure a second magnetic field that is induced by the scleral
coil \cite{reulen_ieee_82}. In another project Roberts et al. suggested to use a resonant circuit 
to read the value of the current inside the scleral coil\cite{dale_etra_08}. A surface-mount device (SMD) capacitor is embedded
inside the contact lens and forms with the scleral coil an LC circuit that generates an oscillating signal proportional to the amplitude of the induced signal.  
These improvements make the SSC approach less invasive, but induce several problems such as a lower signal-to-noise ratio and the need of additional delicate detectors, making the systems more complex and less precise and reliable.

\section{Setup}
\label{sec:setup}

The magnetic tracker used in our experiment is extensively described in two previous papers \cite{biancalana_instrHW_21, biancalana_instrSW_21} and is magnetostatic in nature. We discuss here its potential to perform eye tracking. \correction{Methodologies based on tracking magnetic dipoles have been proposed for other medical purposes such as kinematics evaluations \cite{juce_sens_22}, prosthesis control \cite{tarantino_ieee_20}, endoscopy \cite{than_ieee_12, meng_ieee_21}, surgery \cite{dinatali_ieee_13},  while the application to eye tracking is innovative, and comes with new challenges in terms of the required speed and accuracy.  
In view of eye-tracking application, we test our device} making use of a very simple head+eye simulator that reproduces the basic conditions of eye-tracking measurements and enables qualitative but significant comparisons between the actual movements and those retrieved by the tracker.

\begin{figure}[ht]
\centering%

\subfigure[{Contact lens with magnet}\label{fig:lens_and_magnet}]%
{\includegraphics[width=0.49 \columnwidth ]{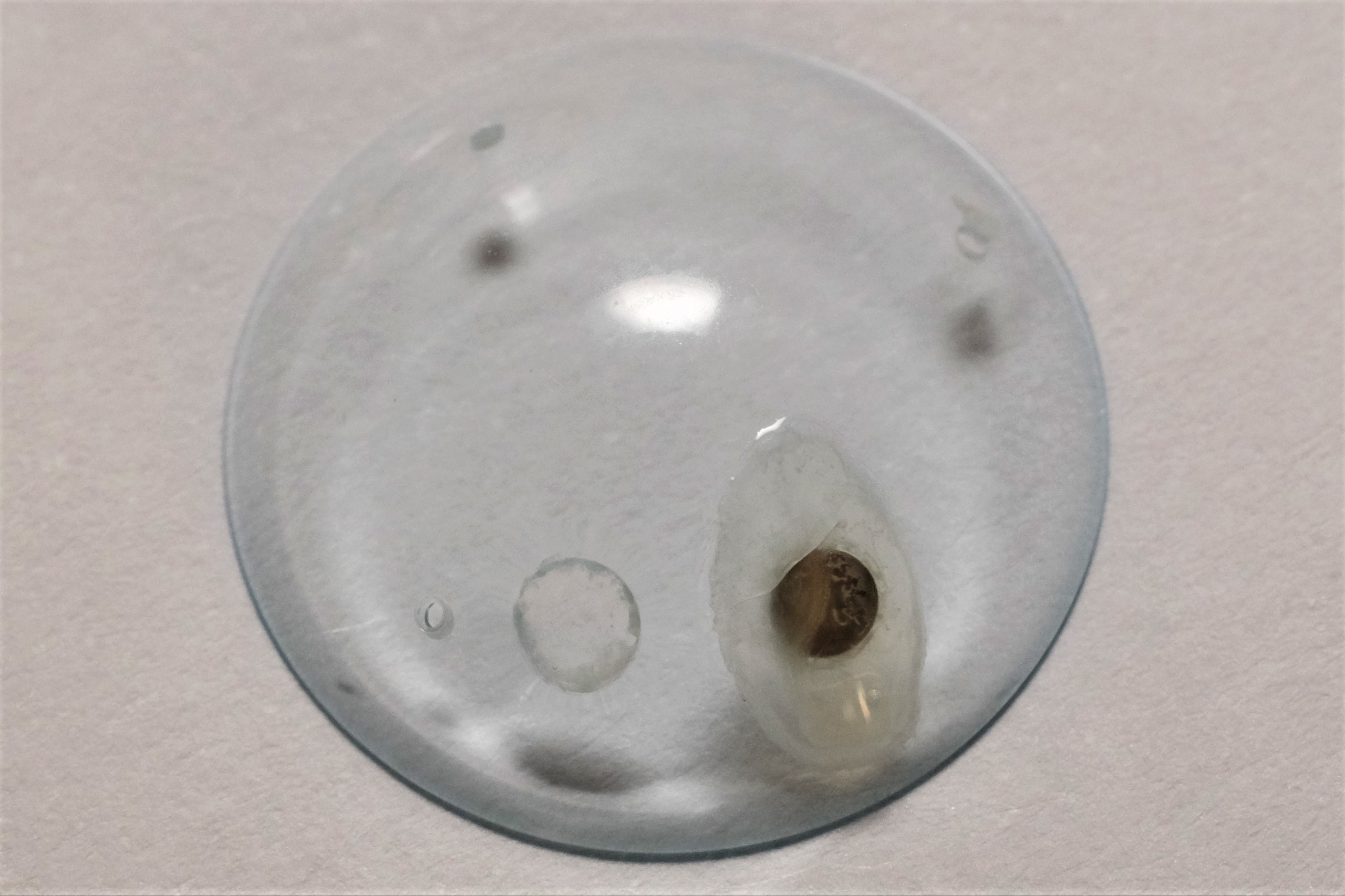}}
\subfigure[{Worn lens}\label{fig:eye_and_lens}]%
{\includegraphics[width=0.49 \columnwidth ]{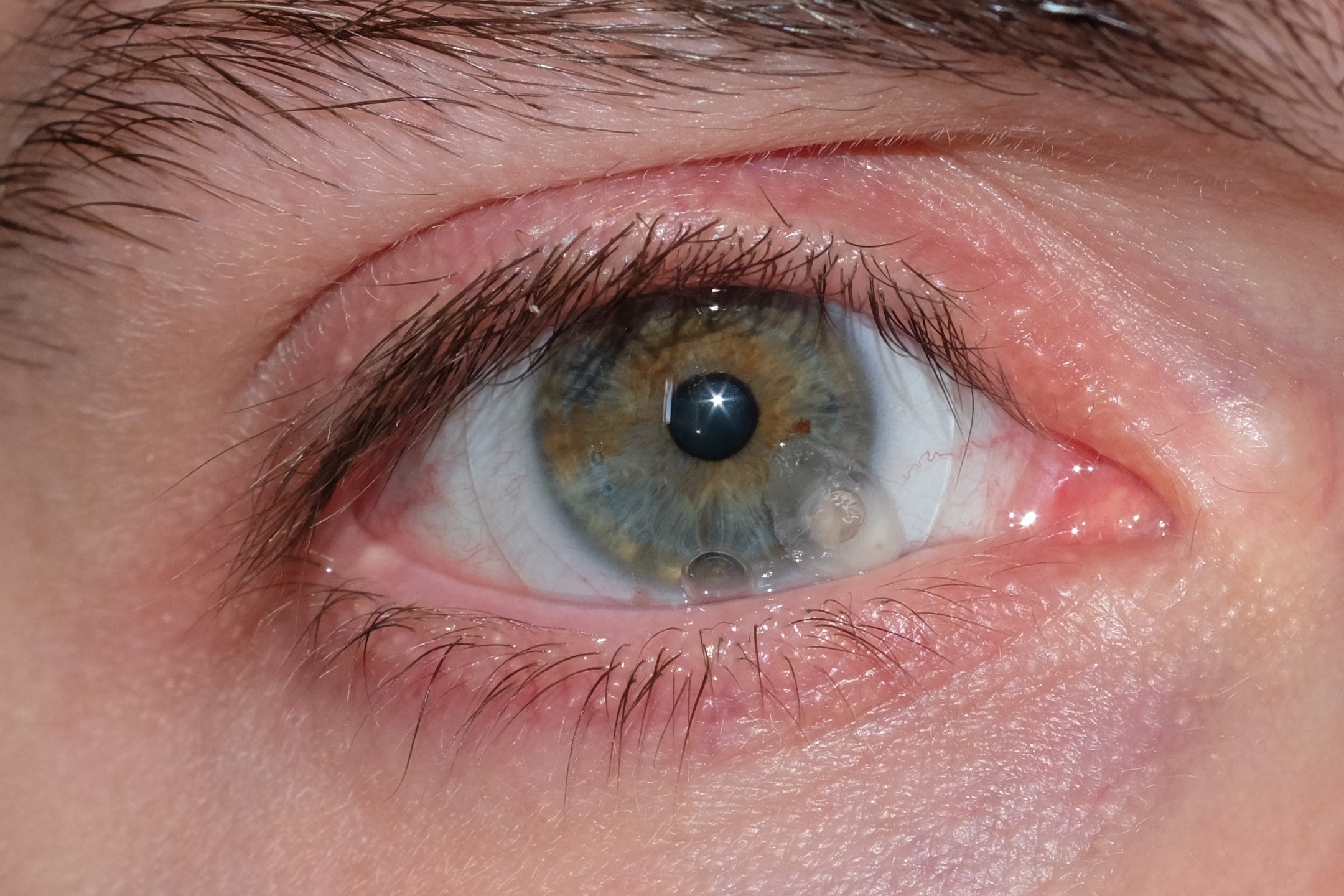}}

\subfigure[{Worn tracker}\label{fig:tracker_and_eye}]%
{\includegraphics[width=0.49 \columnwidth ]{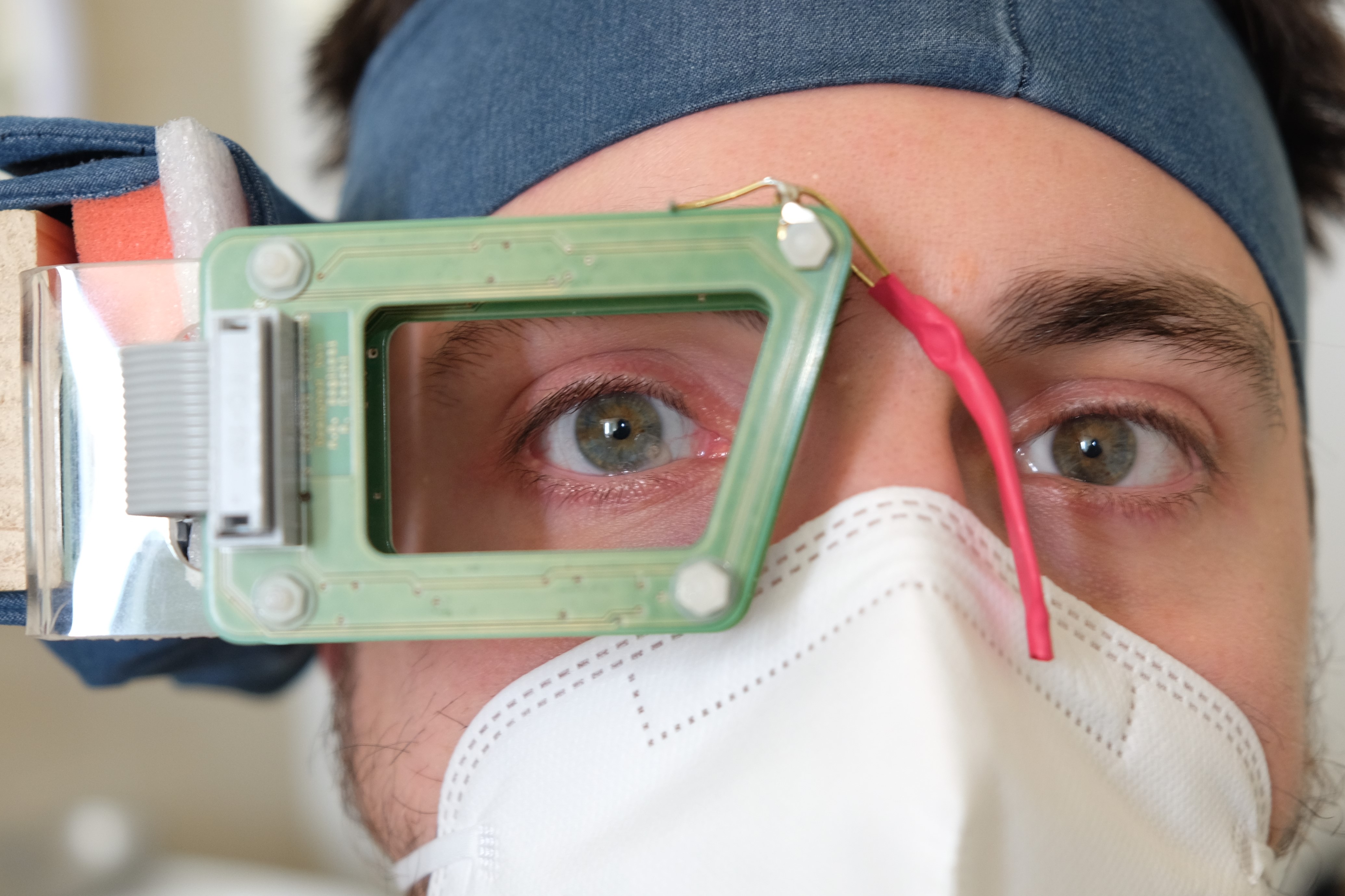}}
\subfigure[{General view}\label{fig:mario_and_tracker}]%
{\includegraphics[width=0.49 \columnwidth ]{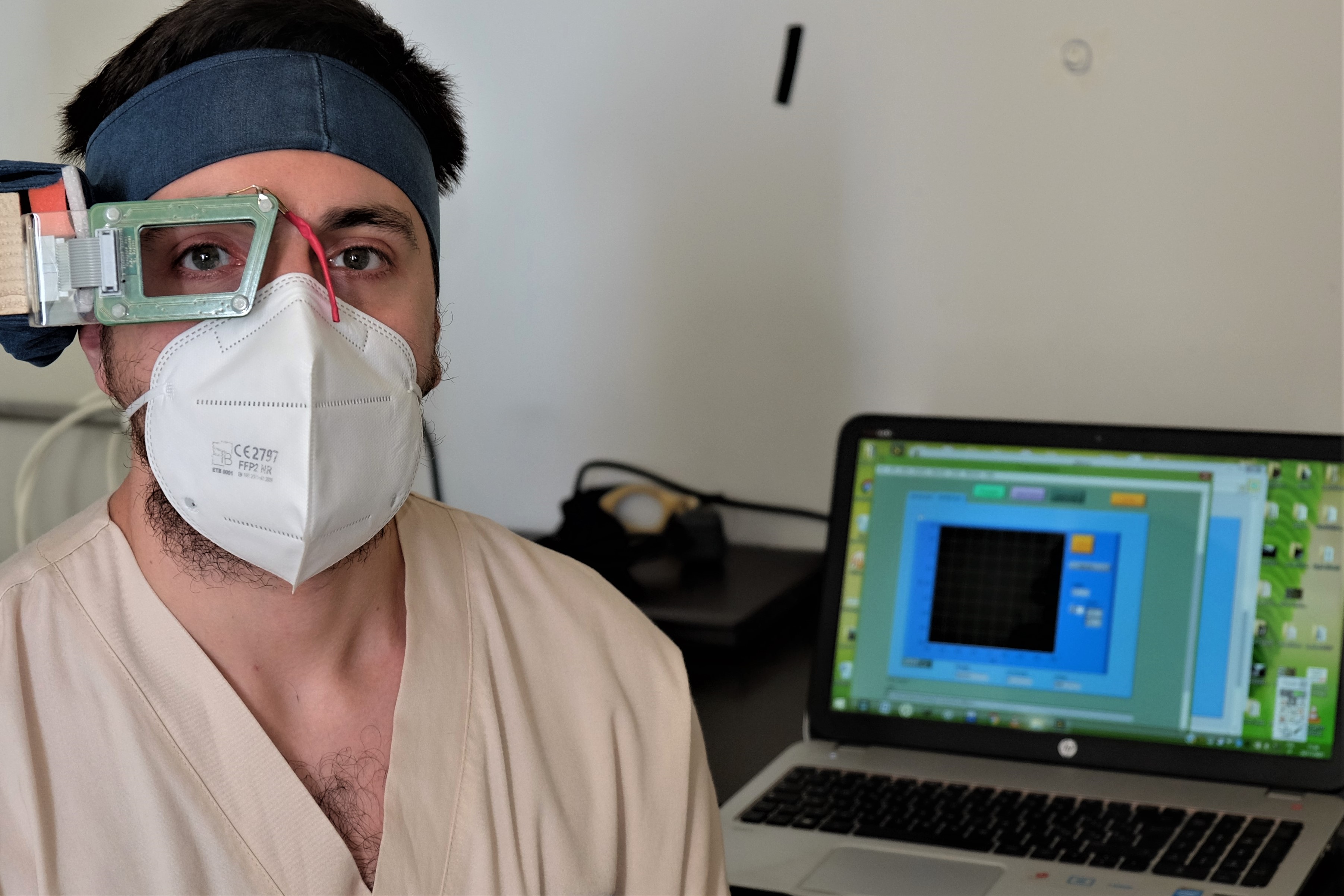}}

\caption{Details of the magnetic tracker under operating conditions: (a) contact lens with an embedded magnet; (b) eye wearing the lens; (c)  worn tracker; (d) general view.} \label{fig:mario}
\end{figure}

In its final application  (real eye tracking), our system will use a magnet embedded inside a contact lens to generate a  field in the eye proximity (see Fig.\ref{fig:mario}. Several 
magnetoresistors  placed in diverse locations in that proximity  detect the dipolar field superimposed to the ambient field, the latter and the source of the dipolar field (i.e. strength, position, and orientation of the magnet) are then numerically evaluated. 

This magnetic eye-tracker is more comfortable 
than the scleral coil, because the eye-sensor connection is inherently wireless. No external field generators are used, which facilitates the head movements and reduces the intrusivity level. The light-weight and compact-size of the sensor array is comparable to instrumentation used in VOG and IR-cam cases, with a similarly low invasivity level.

Compared to VOG and IR-cam, the proposed magnetic eye tracker has a simpler and cheaper construction. Differing from EOG, its operation is not negatively affected by eye-blinking or facial muscles actuation.

\correction{In synthesis, the proposed instrumentation aims to compete with SSC in terms of precision, while offering an  invasivity at level intermediate between  SSC and other techniques (IROG, EOG, VOG), surpassing the latter in terms of robustness, precision, practicality, and cost effectiveness.}

The data analysis produces information about both the position/orientation of the eye with respect to the sensors, and orientation of the latter with respect to the ambient field, so that both eye and head movements are tracked at once. They can be combined (to infer the absolute gaze orientation) or compared (to investigate eye-head correlated movements)\correction{, which is of great importance in some applications \cite{woele_sens_20} and constitutes an added value of the proposed setup. It is worth noting that in several magnetic-tracking applications, the presence of environmental field constitutes a problem because of its interference with the measurement. In contrast, we use an approach where both magnet and external field are accurately analyzed, and both contribute with useful information for such combined eye-head tracking. }

\begin{figure}
\centering
\includegraphics[width=0.48 \columnwidth]{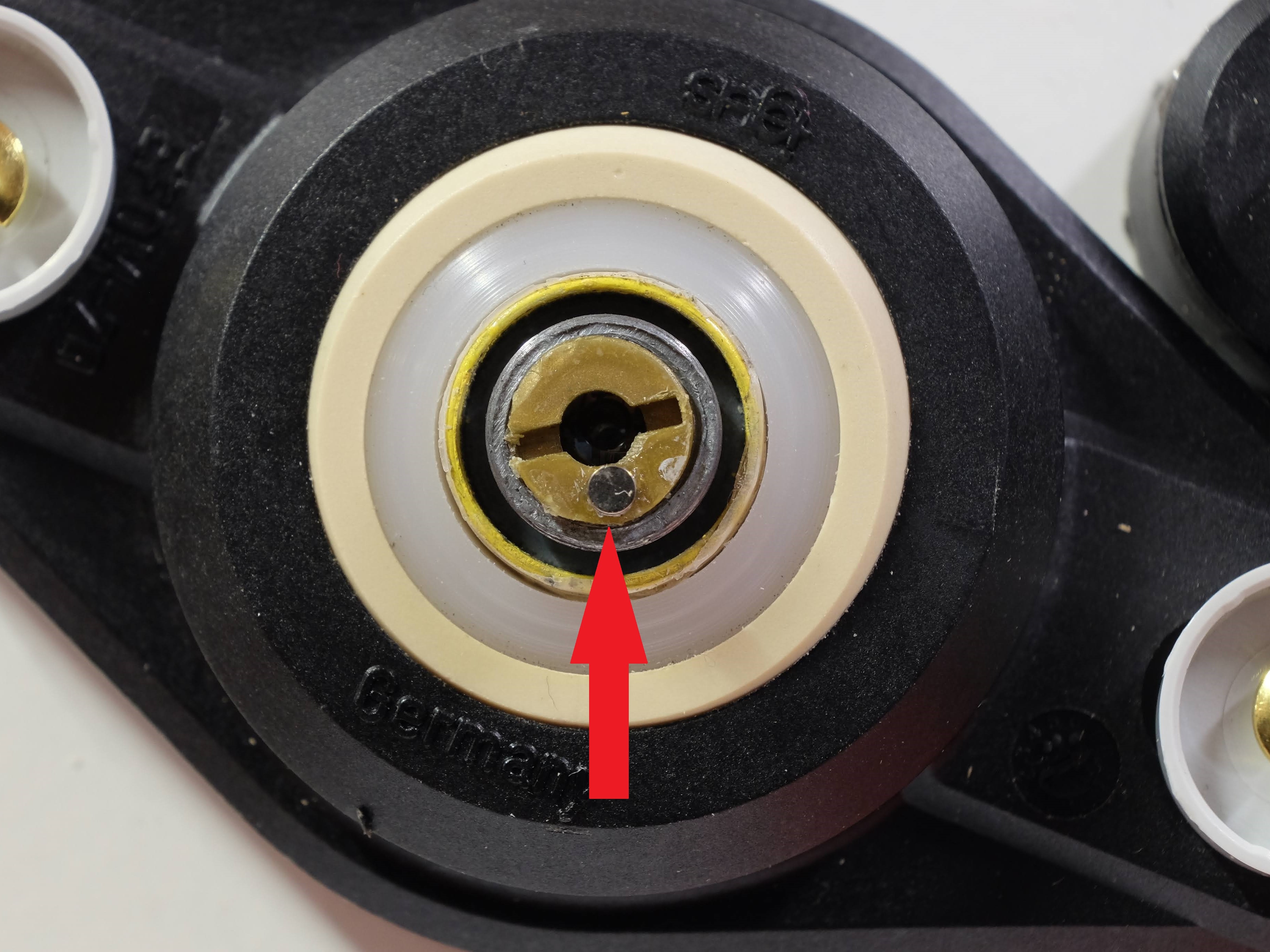}
\includegraphics[width=0.48 \columnwidth]{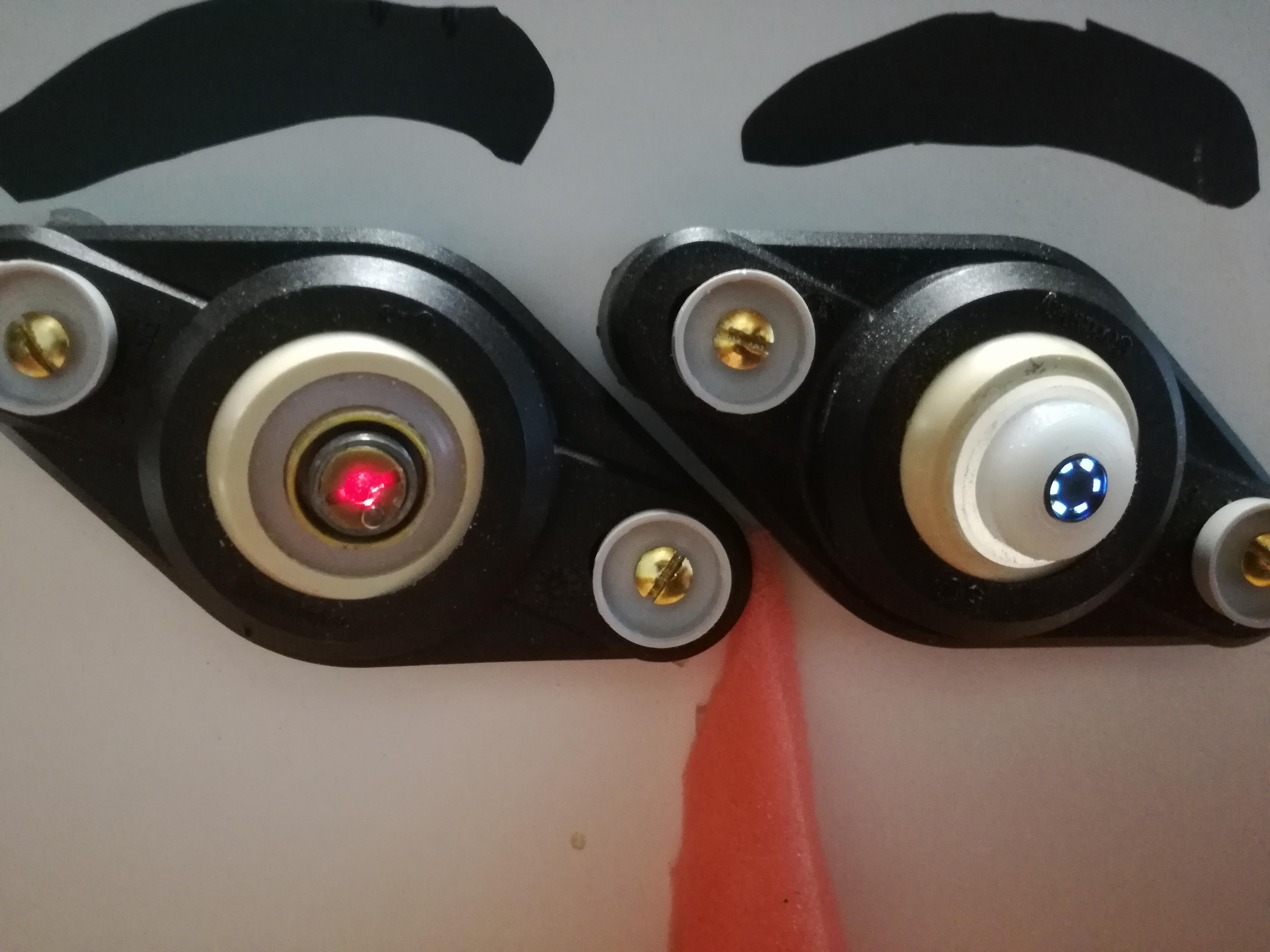}
\caption{\label{fig:simulatore} Two simultaneously actuated spherical bearings (25~mm in diam) host a laser pointer and a microcamera, respectively. A magnet (red arrow in the close-up image at left) is applied in the proximity of the laser output window and is localized by the tracking system. The bearings reproduce motion of the eyes and are mounted on a movable frame that represents the head.
\href{https://photos.app.goo.gl/rq9En5AyyzitDFRR7}{*link VIDEO1*} \correction{(Multimedia view)}
}
\end{figure}

The head-eye simulator used to test the tracker (see Fig.~\ref{fig:simulatore}) \correction{(Multimedia view)} is based on a couple of connected spherical bearings (25~mm in diameter) that are hand-actuated and host a laser pointer and a microcamera, respectively. The mechanical connection lets the bearings turn horizontally and vertically, while maintaining their parallelism.
This system provides reliable visualization of the gaze direction (set by eye and head orientations) via the position of the projected pointer spot, while the micro-camera applied to the second eye reproduces the viewed area.

A small magnet (2~mm diameter 0.5~mm thickness) is pasted nearby the exit window of the laser pointer: in other terms the laser window simulates the  pupil of the tracked eye. The gaze orientation can be visualized in terms of the laser spot position, as well as in terms of the central zone of the area framed by the microcamera.
In the next Section, we are presenting the derivation of the gaze orientation from the tracker output, which enables a comparison of the tracker output with the observed displacements of the laser spot projected on a screen and/or with the scene recorded by the microcamera.

\section{Gaze estimation}
\label{sec:calculus}
As  discussed in Refs.\onlinecite{biancalana_instrHW_21, biancalana_instrSW_21}, the magnetic tracker measures the magnetic field  in several assigned positions around the eye, which is marked with a small magnet (dipole) inserted in a contact lens, and uses a best-fit procedure to infer the position and the orientation of the dipole, as well as the components of a homogeneous field superimposed to the dipolar field. The magnetic sensors are triaxial magnetoresisitive detectors, whose digitized output is numerically converted into magnetic units with the help of an accurate calibration procedure.  
The best-fit procedure inputs the sensor position and the field measurements and outputs position and orientation of the dipole, and the environmental magnetic field.
The best fit procedure uses a Levemberg-Marquardt minimization algorithm, and can elaborate --running in a Intel-i7 2.40~GHz PC personal computer-- one hundred data sets per second, which is the maximum acquisition rate currently allowed by the microcontroller that interfaces the sensors to the computer. According to the sensor specifications \cite{isentek8308}, this rate can be doubled, and even extended to 1kSa/s with an updated version of the chip\cite{isentek8308A}. 
The size of the permanent magnet is selected in such way to produce a magnetic field comparable with the Earth's magnetic field ($B\sim 40\mu$T) at its typical (centimetric) distance from the sensors. Thus every sensor measures a superposition of those two magnetic fields, in such a condition that none of them contributes predominantly/negligibly on the measured quantities. A magnetic dipole $\vec{m}$ placed in the position $\vec{r}=(x,y,z)$ makes the $i^{\textrm{th}}$ sensor (located in $\vec r_i$) measure a magnetic field
\begin{equation}
    \vec{B}(\vec{r},\vec{m})=\frac{\mu_0}{4\pi} \left(3\frac{(\vec{m}\cdot \vec{R_i}) \vec{R_i}}{|\vec R_i|^5}-\frac{\vec{m}}{|\vec R_i|^3} \right) + \vec{B_{g}} \label{eq:magnetic_measure}
\end{equation}
where $\mu_0$ represents the vacuum permeability and $\vec{B_{g}}$ is the environmental field, assumed homogeneous, and $\vec R_i= \vec r _i -\vec r $ is the position of the sensor relative to the dipole.
In the Eq.~\ref{eq:magnetic_measure}, $\vec{r}$, $\vec{m}$ and $\vec{B_{g}}$ are unknowns, for a total of $n_p=9$ parameters, as each of these three vectors has three independent components. The mentioned best-fit procedure
enables the determination of these $n_p$ parameters starting from $K$ 
independent simultaneous measurements of $\vec B$ components, with the obvious requirement $K 
\ge n_p$. A large $K$ value helps the procedure produce more reliable and precise 
estimations, particularly under the condition that the relative positions $\vec R_i$ are quite different from each other. In our 
case, using eight triaxial sensors, we have $K=24$. As previously pointed out \cite{biancalana_instrSW_21}, it can be advantageous (in terms of accuracy rather than in terms of computation velocity) to reduce $n_p$ on  the basis of the constraint that $|\vec m|$ is constant, so that only two angular parameters must be determined to identify the dipole orientation. 
In contrast, possible weak inhomogeneities of the environmental field, despite being negligible over the sensor-array baseline, may produce non-negligible variations of the $B_g$ modulus when the array is freely displaced. As a consequence, it is definitely inopportune to further reducing $n_p$ on the basis of a $B_g$=constant constraint. In any case, the tracking procedure outputs a set of 9 parameters, namely the array $(\vec r, \vec B, \vec m)= (x, y, z, B_x, B_y, B_z, m_x, m_y, m_z)$.

\begin{figure}
\includegraphics[width=0.7 \columnwidth]{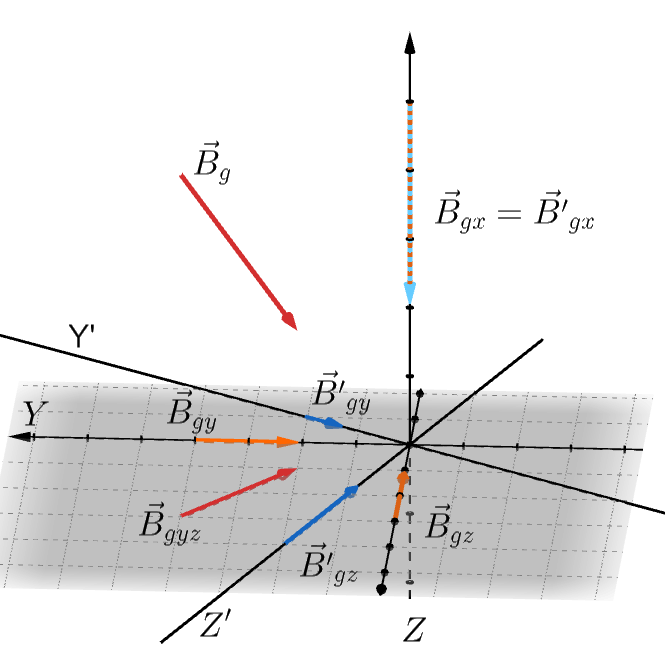}
\caption{\label{fig:magnetic_field} Arbitrarily oriented 
environmental magnetic field $\vec{B_g}$ (red arrow) in two Cartesian co-ordinate systems that are rotated around the $x$ axis with respect to each other. The rotation causes a variation of the $y$ and $z$ components (from orange to blue), while the $x$ component (blue-orange dotted arrow) keeps constant. }
\end{figure}

The head orientation can be inferred from the  $(B_x, B_y, B_z)$ components. We are mainly interested in small (< 1 rad) rotations around a vertical ($x$) and around a horizontal ($y$) axis (see Fig.\ref{fig:head_rotation}), which, with the notation defined in Appendix \ref{app:angleandaxis}  corresponds to using $\hat u= (0,1,0)$ and $\hat u= (1,0,0)$, respectively.
while in these tests, we do not consider rotations around the $z$ axis, which would correspond to lateral inclinations of the head. Similarly (see below) we are not analyzing eye rotations around $z$, i.e. eye torsional motion.

\begin{figure}[ht]
\centering%
\subfigure[{Head rotation \href{https://photos.app.goo.gl/qCRNsCfsPjsBszt56}{*link VIDEO2*} \correction{(Multimedia view)}
}\label{fig:head_rotation}]%
{
\includegraphics[width=0.49 \columnwidth ]{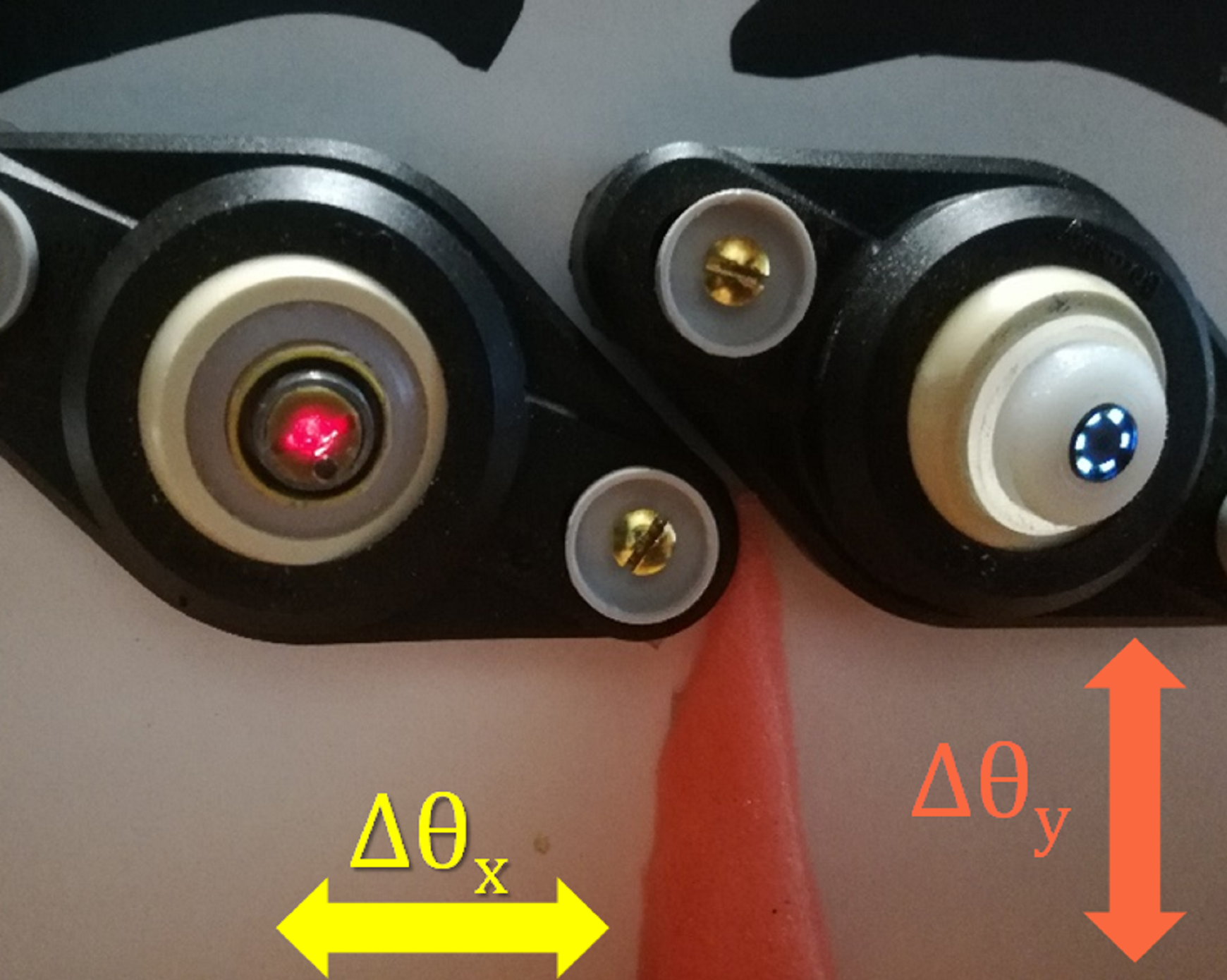}}
\subfigure[{Eye rotation \href{https://photos.app.goo.gl/JCzDvDRzkCwkQZD96}{*link VIDEO3*} \correction{(Multimedia view)}} \label{fig:eye_rotation}]%
{\includegraphics[width=0.49 \columnwidth]{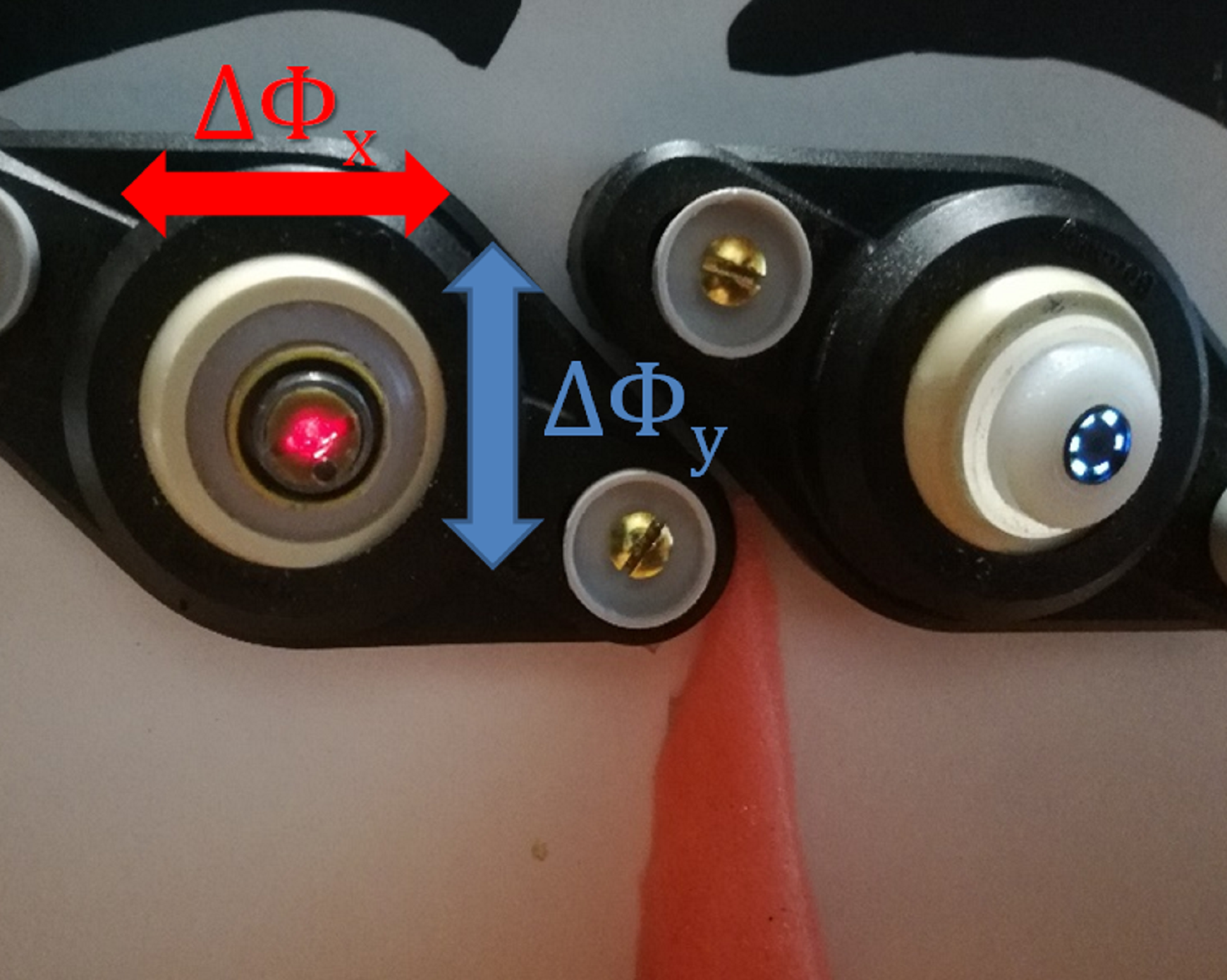}}
\subfigure[{Head and eye rotation \href{https://photos.app.goo.gl/8ZAYgRSWbKmMZS628}{*link VIDEO4*} \correction{(Multimedia view)} }\label{fig:HE_rotation}]%
{\includegraphics[width=0.49 \columnwidth]{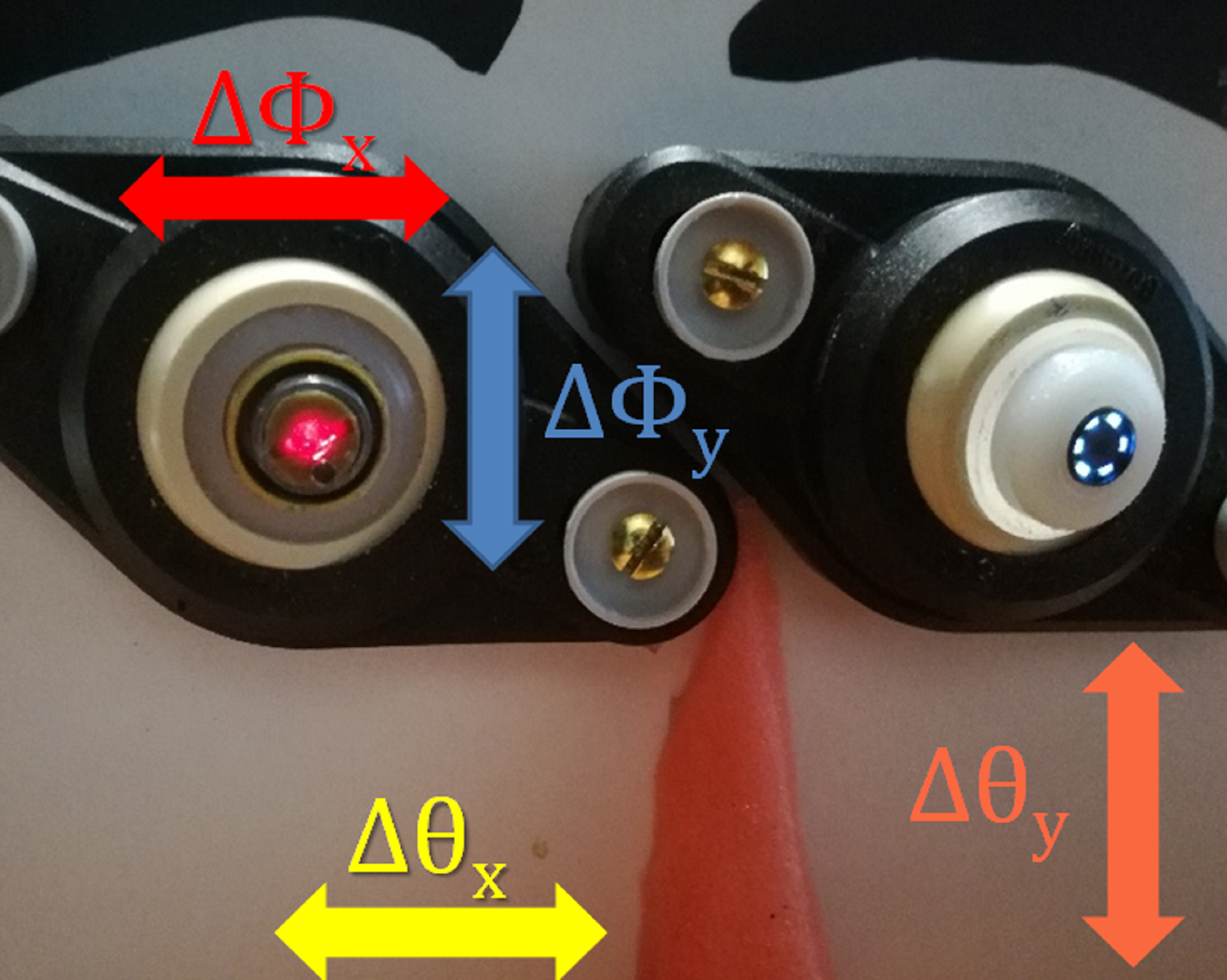}}
\subfigure[{Head and eye angles} \label{fig:angledefinition}]%
{\includegraphics[width=0.49 \columnwidth]{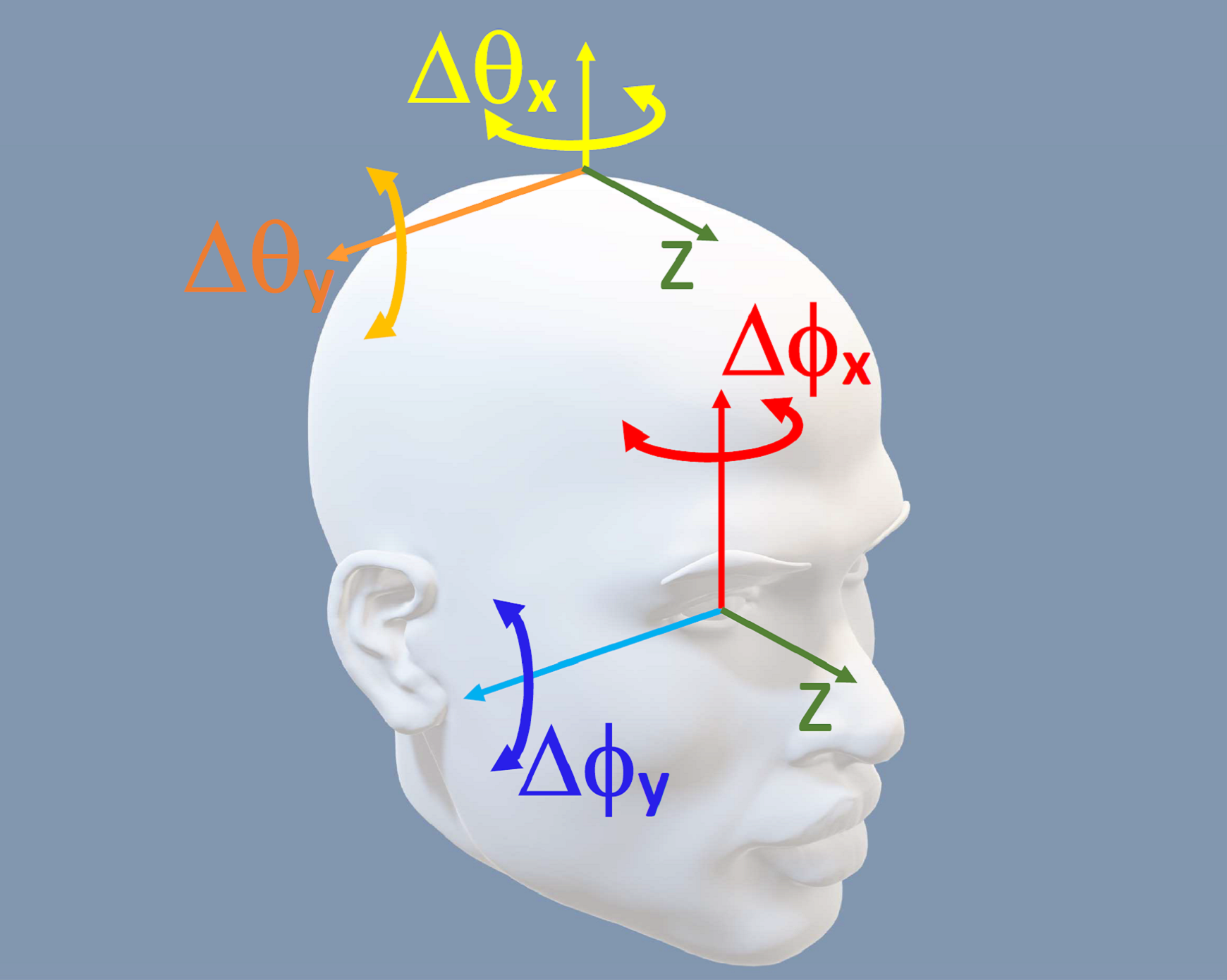}}

\caption{Rotation angles of the head (a) and of the eye (b). Head lateral rotation ($\Delta \theta_x$)  and lean-bow rotation ($\Delta \theta_y$) can be inferred from the apparent rotation of the environmental field. Eye rotations can be  inferred from dipole orientation ($\Delta \phi_{x,y}^{(m)}$) or --alternatively-- from magnet displacements ($\Delta \phi_{x,y}^{(p)}$). It is possible to visualize a combination of head and eye rotations (c). The multimedia video shows how --in this latter case--  the zero-angle gaze is recovered, when the eye rotation compensates the head rotation. The  definition of angles is summarized in (d).} \label{fig:MovementsHE}
\end{figure}

The evaluation of $\theta_x$ (head rotation around the vertical direction) would be unreliable in the case that both $B_y$ and $B_z$ vanishes, i.e. if the environmental field is vertical: this is not the case at non-polar latitudes. The evaluation of $\theta_y$ (head rotation around a transverse horizontal direction) would be unreliable in the case that both $B_x$ and $B_z$ vanish, i.e. if the environmental field is horizontal and transverse with respect to the front direction: also this problem is not expected to occur at our (non equatorial) latitudes, where the vertical component ($B_x$) is non-zero. 

It is worth noting that the non-commutative nature of finite-angle rotations makes the two estimations of $\Delta\theta_x$ and $\Delta\theta_y$ incompatible with each other, particularly when large rotations are considered. However a composite estimation for arbitrary head movement is feasible with good accuracy, provided that rotations amount to a small angle. Analogous limitations applies to the estimation of eye orientation discussed here below. 
In contrast, larger finite rotations around an assigned direction facilitate an \textit{a posteriori} identification of that axis, on the basis of what is reported in the Appendix \ref{app:angleandaxis}.
In addition, for the sake of simplicity we use the same symbols for head a eye rotation axes, however it should be noted that the co-ordinate system used to describe the eye orientation moves together with the head. E.g. the $x$ ($y$) axis is vertical (horizontal and transverse) for the both the head and the eye only when both head and eye are front oriented (zero-angle gaze), as in the configuration represented in Fig.\ref{fig:angledefinition}.

The orientation of the eye with respect to the sensor array (and hence to the head)  can be inferred either from the $(x, y, z)$ or from the $(m_x, m_y, m_z)$ analysis, see Fig.\ref{fig:eye_rotation}. We will refer to the two independent estimates with the symbols $\phi^{(p)}$ and $\phi^{(m)}$, respectively. In the case here presented (with $\vec m$ oriented along $z$ in front-glance condition), these alternative estimates of $\phi$ offer a redundancy that can be profitably used to improve the accuracy or to point out unexpected system faults.
Using the spatial co-ordinates, lets roughly estimate the eye orientation on the basis of some reasonable (but approximate) assumptions. Using for the eye radius $R_{\mathrm{eye}}$  its typical value (about 12~mm) and assuming that the center of the eye-ball is displaced only along the $z$ direction with respect to the magnet when the eye glance is front-oriented, the rotations of the eye (with respect to the head) around the vertical and horizontal axes can be estimated as 
\begin{equation}
\Delta \phi_x^{(p)} = \arcsin \left( \frac{y-y_0}{R_{\mathrm{eye}}} \right ),
\label{eq:occhioxP}
\end{equation}
and 
\begin{equation}
\Delta \phi_y^{(p)}= \arcsin \left ( \frac{x-x_0}{R_{\mathrm{eye}}} \right ),
\label{eq:occhioyP}
\end{equation}
respectively, being  $(x_0, y_0)$ the measured co-ordinates when the glance is front-oriented.

The magnet orientation provided by the $(m_x, m_y, m_z)$ parameters determined by the best fit enables an alternative estimation that does not require assumptions on the eye center and radius. In particular, the eye rotations around the $x$ and $y$ directions, $\Delta \phi_x^{(m)}$ and $\Delta \phi_y^{(m)}$, can be estimated with the same approach previously described for the head (Appendix \ref{app:angleandaxis}).

The magnets used in the present experiment are thin disks with axial magnetization, thus the dipole will be (nearly) oriented along $z$ in a front-glance condition in real application, when the disk is inserted in a contact lens. As a consequence,  $\Delta \phi_x^{(m)}$ and $\Delta \phi_y^{(m)}$ are reliably determined while, torsional motion (i.e. eye rotation around the $z$ axis) could not be accurately detected: to this aim, diametrically magnetized disks should be used, instead. In this case the above mentioned redundancy would turn to an unprecedented feature, because the independent measures of $\Delta \phi^{(m)}$ and $\Delta \phi^{(p)}$ will provide complete (3D) information about the eye orientation. For example, if in front-glance condition $\vec m$ is oriented along $x$, beside the quantities 
$\Delta \phi_x^{(p)}, \Delta \phi_y^{(p)}$ estimated with Eqs.\ref{eq:occhioxP}, \ref{eq:occhioyP}, the angles $ \Delta \phi_y^{(m)}, \Delta \phi_z^{(m)}$ would be accessible, leading to a 3D angular tracking with redundant $\Delta \phi_y$. In summary, even if not accessible with the axially magnetized magnets discussed in this work, the apparatus may be used to analyze eye torsion, which is a relevant added value in important medical applications \cite{zee_jov_15}.

\section{results}
\label{sec:results}

As discussed in Sec.\ref{sec:calculus}, the information provided by the tracker lets retrieve  information about head and eye rotation independently. The  eye and the head rotation angles can be visualized independently versus time, one versus another (i.e. in plots $\Delta\theta_x $ vs. $ \Delta\theta_y$, Fig.~\ref{fig:head_rotation} and $\Delta\phi_x$ vs. $\Delta\phi_y$, Fig.~\ref{fig:eye_rotation}) or combined together to reproduce the gaze orientation, plotting $\Delta\theta_x+\Delta\phi_x $ vs. $ \Delta\theta_y+\Delta\phi_y$, as shown in  Fig.~\ref{fig:HE_rotation}.

\begin{figure}[ht]
\centering%
\subfigure[{Screen setup \href{https://photos.app.goo.gl/yEUwEWjBfEmcMgee7}{*link VIDEO5*} \correction{(Multimedia view)}}\label{fig:screenspot}]%
{\includegraphics[ width=0.47 \columnwidth]{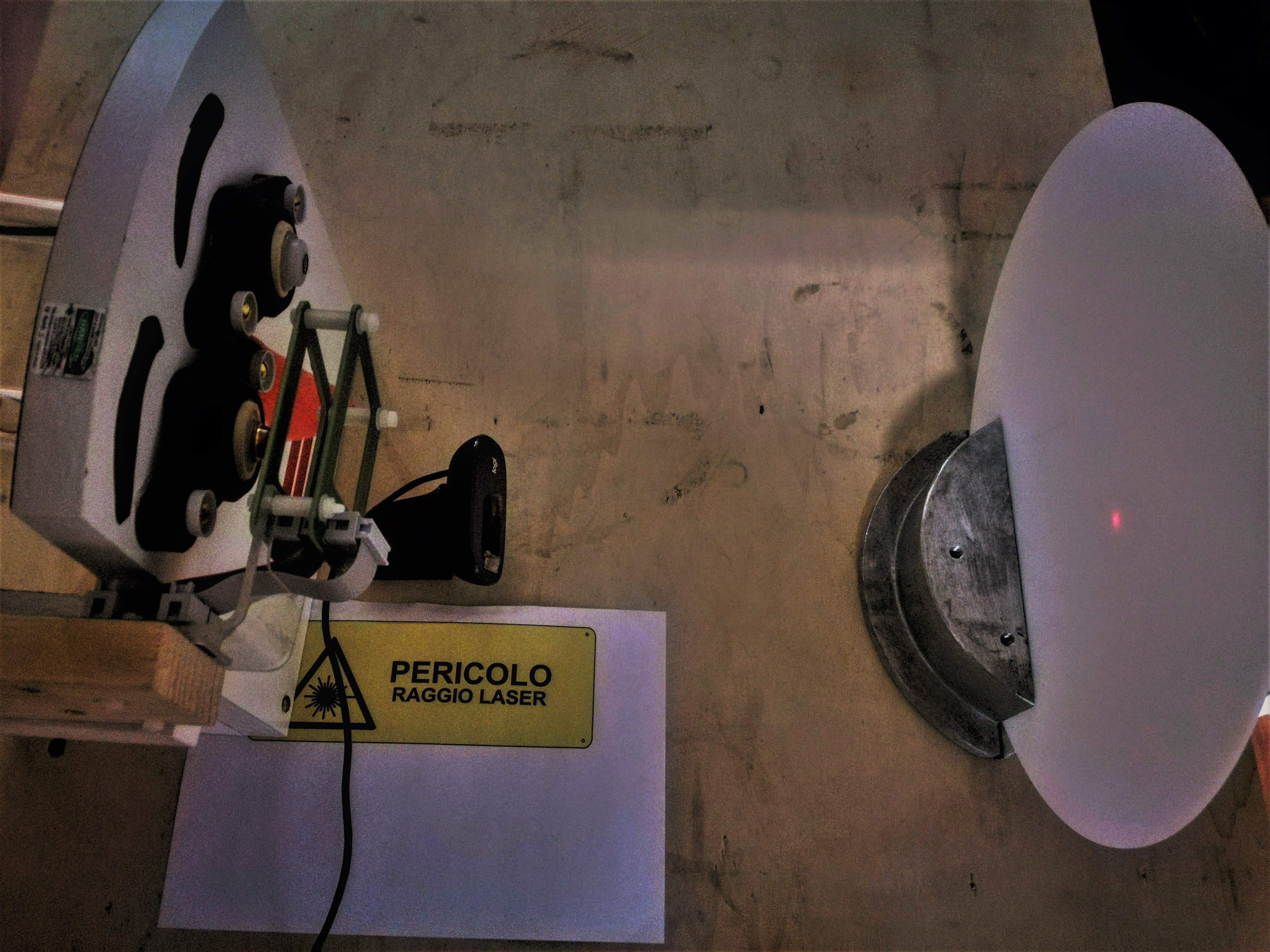}}
\subfigure[{Framed scene \href{https://photos.app.goo.gl/wtRdfSpfRLrvDH2H6}{*link VIDEO6*} \correction{(Multimedia view)}} \label{fig:microcamera}]%
{\includegraphics[width=0.47 \columnwidth]{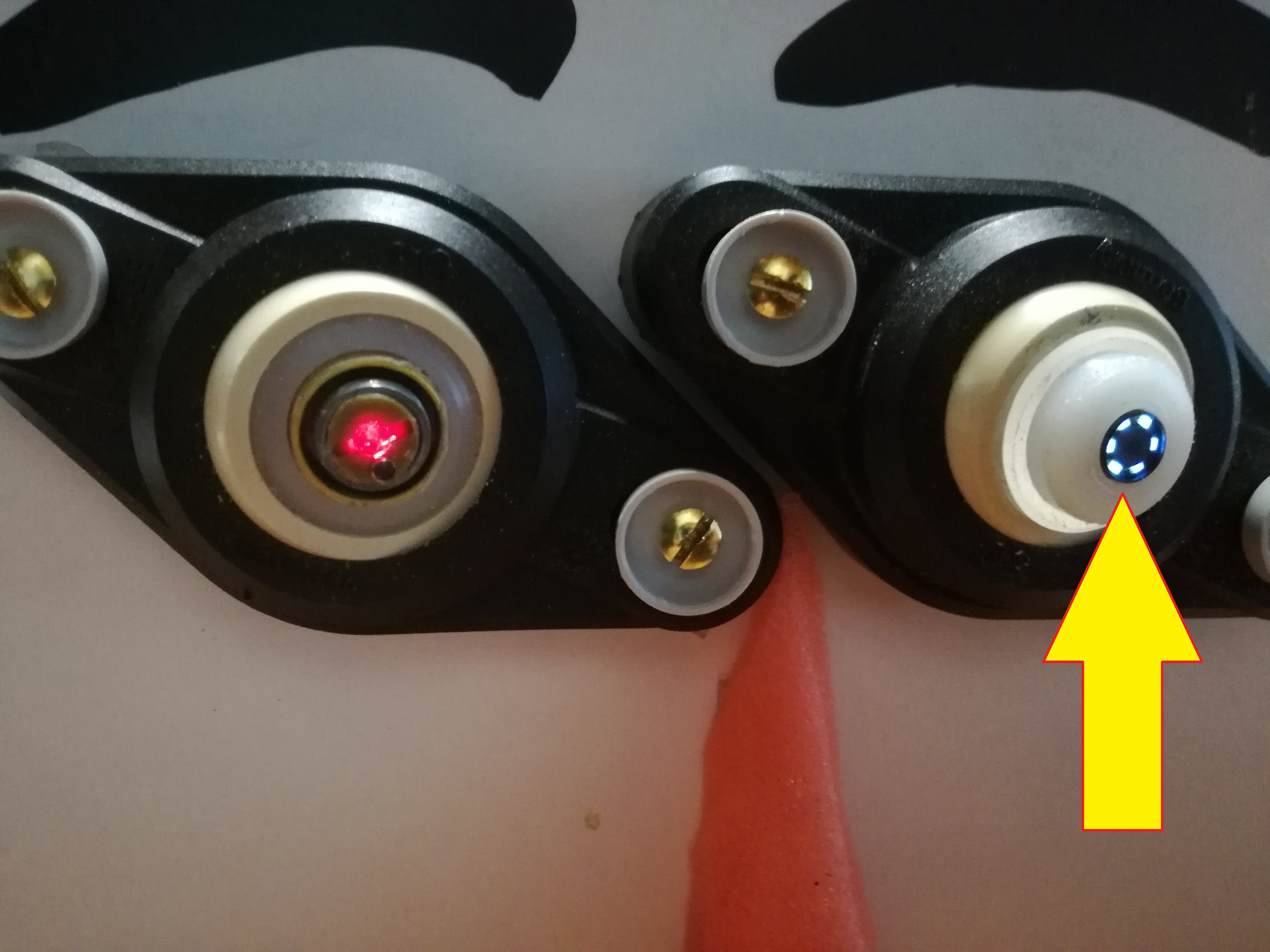}}
\caption{The red spot of the laser shows the intercept of gaze direction with the projection screen (a). The multimedia video demonstrates the tracker performance comparing the movements of the laser spot on the screen and the reconstructed ($\Delta \theta_x + \Delta \phi_x, \Delta \theta_y + \Delta \phi_y$) position of the magnet. Alternatively, (b), the gaze direction can be inferred from the area framed by the microcamera}\label{fig:MovementsSpot}
\end{figure}

It is significant to compare the tracker outputs with head, eye, or head+eye motion applied to the simulator. The latter can be visualized as direct images (movies) of the
simulator  as shown in Fig.~\ref{fig:MovementsHE} and related multimedia, or  looking at the laser spot projected on a screen (or, equivalently, to the scene framed by the videocamera), as shown in (Fig.~\ref{fig:MovementsSpot}).

Beside the qualitative analyses reported with the multimedia related to Figs.~\ref{fig:MovementsHE} and \ref{fig:MovementsSpot}, some quantitative estimations of the angle uncertainties are performed. Additional information about tracking uncertainties, particularly for the dipole position,  is provided in ref. \onlinecite{biancalana_instrSW_21}. 

The random error on the angles is quantified in terms of standard deviation in large samples of angle estimations obtained from measurements in steady conditions (fixed head and eye orientation).

Environmental magnetic disturbances and detector fluctuations that may originate, e.g., from thermal or digitization noise produce small variations on the measured components of $\vec B$, which are then transferred on the tracking parameters by the best-fit algorithm. The input noise level can be reduced by performing a data average at the detection level and/or acquiring oversampled data. 
The detectors can be set to perform such data averaging (at expenses of lowering the effective bandwidth) and oversampling.

The estimated random errors on the eye and head angles are reported in Tab.\ref{tab:std1} for different settings of data averaging and a 4$\times$ oversampling rate. The error on head angles $\sigma_{\theta}$ may vary with the absolute orientation of the head, due to the more or less favorable conditions (discussed in Sec.\ref{sec:calculus}). In this instance, the worst case for the head uncertainty is $\sigma_{\theta} < 1$~mrad. The random error for the eye orientation angles is about four times larger, and has comparable amplitudes if it is estimated on the basis of the magnet position (Eqs.~\ref{eq:occhioxP}, \ref{eq:occhioyP} ) or of the magnet orientation.

\begin{table}[ht]
\begin{center}
\begin{tabular}{c|c|c||c|c||c|c||}
\cline{2-7}
\multicolumn{1}{c||}{} & 
\multicolumn{2}{c||}{$\sigma_{\theta_x}$} & 
\multicolumn{2}{c||}{$\sigma_{\phi_x^{(m)}}$} & 
\multicolumn{2}{c||}{$\sigma_{\phi_x^{(p)}}$}\\ 
\cline{2-7}

\multicolumn{1}{c||}{}  & mrad & $\degree$ & mrad  & $\degree$ & mrad & $\degree$ \\
\cline{1-7}
\multicolumn{1}{|c||}{Avg 1} & 0.86 & 0.051 & 3.4 & 0.19 & 3.6 & 0.21\\
\hline
\multicolumn{1}{|c||}{Avg 2} & 0.59 & 0.034 & 2.1 & 0.12 & 2.4 & 0.14\\
\hline
\multicolumn{1}{|c||}{Avg 4} & 0.43 & 0.025 & 1.3 & 0.074 & 1.7 & 0.097\\
\hline
\multicolumn{1}{|c||}{Avg 8} & 0.18 & 0.010 & 1.1 & 0.063 & 1.5 & 0.086\\
\hline
\end{tabular}

\caption{Uncertainties on the estimated angles due to measurement fluctuations under conditions of static eye and head orientation, expressed in mrad and in degrees. Large samples of repeated measurements and subsequent best-fit estimations are used to compute their standard deviation. At the expense of a reduced bandwidth, the sensors may return  values averaged over 2, 4, 8 samples, with progressively lower noise levels on the magnetometric data and consequently more stable angle estimations. }\label{tab:std1}

\end{center}
\end{table}

The estimate of the eye orientation ($\Delta \phi_{x,y}$) is affected --to a small extent-- by variations of the head orientation, i.e. there is a residual cross-talk among the two quantities. Evaluating the eye-angle random error $\sigma_\phi$ over measurement sets recorded while the eye is fixed with respect to the head, but the head rotates in the environmental field, causes an increase of the standard deviation. Of course, this increase is strongly dependent on the specific degree of  inhomogeneity of the ambient field where the measurement is performed. In our working conditions and when no preliminary averaging is performed, we observe a typical increase by a factor of 5, up to $\sigma_\phi \approx 15$~mrad ($1\degree$).

\section{Conclusion}
\label{sec:conclusion}

We have tested an innovative eye-tracker setup based on measuring the magnetic field produced in the proximity of the eye by means of a set of magnetoresistive detectors.

Several advantages of the proposed approach to  eye-tracking with respect to other available technologies  have been  discussed. In particular, we have pointed out how the system can record torsional eye movements, a peculiarity shared  with some implementations of the magnetic search coil technique and barely achievable with other methodologies \cite{zee_jov_15}. With respect to the search-coil apparatuses, the  system here described is more compact and portable. Apart from the need of using a contact lens, the proposed technique  is neither invasive nor intrusive, and enables the construction of wearable devices more precise and cost-effective than other low-invasivity competing technologies. In addition, the system enables simultaneous determination of environmental field and position-orientation of the dipolar source. This feature facilitates the task of comparing head and eye movements, which is crucial in some medical diagnostics, particularly in those for vestibular diseases.

The system performance has been analyzed by qualitative but significant comparisons of eye and head motion with the corresponding tracking results.
A simple simulation hardware making use of a laser pointer and a microcamera made possible to operate the tracker under realistic conditions, while simultaneously producing objective records of the actual gaze orientation, which enables direct visual comparison of actual and tracked gaze direction.

\begin{acknowledgments}
L.B, V.B. and G.B. acknowledge the partial support of Regione Toscana, under the project PhAST (FESR 2014 – 2020).
\end{acknowledgments}

\section* {Data Availability Statement}
The data that support the findings of this study are available from the corresponding author upon reasonable request.

\appendix

\section{Angle and rotation axis estimation}
\label{app:angleandaxis}
The measurements described in the main text require an extimation of rotation angle of the measured $\vec m$ and $\vec B$ vectors, the former due to eye rotation with respect to the sensor frame, and the latter due to the apparent rotation of the environmental field that appears in the co-ordinate system of the sensor frame that moves integrally with the head.

The rotation angle of a vector $\vec A$ around a given direction $\hat u$ ($\hat u$ being a unitary vector oriented along the rotation axis) is estimated as follows.

Let $\vec A_i$, with $i=1,2$ be two subsequent measurements of $\vec A$, so that $\vec A_{i, \perp }= \vec A_i -(\vec A_i \cdot \hat u) \hat u $ are their projections on a plane perpendicular to $\hat u$. The rotation angle must be measured on that plane, and is inferred from the projected vectors s' cross product and moduli as
\begin{equation}
    \psi=\arcsin{ \frac{|\vec A_{1, \perp } \times \vec A_{2, \perp }|}{|A_{1, \perp }||A_{2, \perp }|}}
\end{equation}
that produces reliable results, as far as the rotation does not exceed $\pi/2$, which means \textit{always} in the considered application.

While in some applications the rotation axis $\hat u$ is known \textit{a priori}, there are instances where a finite rotation occurs around an unpredictable, but fixed direction. In these cases, provided that a set of at least three measurements \{ $\vec A_i $\} is available, the vector $\hat u$ can be estimated as the normalized average of  quantities $(\vec A_i-\vec A_j) \times (\vec A_{i+k}-\vec A_{j+l})$, with $j>i$, $k,l>0$: in other terms, the rotation axis can be inferred from the cross product of subsequent variations of the measured vector.  With an advantageous simplification,  $\hat u$ can be expressed in terms of the quantity 
\begin{equation}
    \sum_{i=1}^{N-1} \vec A_i \times \vec A_{i+1}+ \vec A_N \times\vec A_1,
\end{equation}
opportunely normalized.

\bibliographystyle{ieeetr}
\bibliography{TrackerBib}

\end{document}